\pdfoutput=1
\documentclass[11pt]{article}
\usepackage[T1]{fontenc}
\usepackage[utf8]{inputenc}
\usepackage{textcomp}
\usepackage{amsmath,amssymb}
\usepackage[margin=1in]{geometry}
\usepackage{newpxtext,newpxmath}
\usepackage{graphicx}
\usepackage{placeins}
\usepackage{xcolor}
\usepackage[super,sort&compress,comma]{natbib}
\usepackage[font=small,labelfont=bf,labelsep=period,justification=justified]{caption}
\usepackage{titlesec}
\titleformat{\section}{\large\bfseries}{\Roman{section}.}{0.6em}{\MakeUppercase}
\titleformat{\subsection}{\normalsize\bfseries}{\Alph{subsection}.}{0.6em}{}

\usepackage{authblk}

\usepackage{xurl}
\usepackage[colorlinks=true,linkcolor=blue!50!black,urlcolor=blue!50!black,citecolor=blue!50!black]{hyperref}
\DeclareUnicodeCharacter{2212}{\ensuremath{-}}
\usepackage{array}\newcolumntype{L}[1]{>{\raggedright\arraybackslash}p{#1}}

\title{\Large\bfseries Diagnosing internal leakage currents through magnetometry of lithium-ion cells
}
\author[1,2]{Shaurya Shukla}
\author[2]{Chengtong Zhang}
\author[3]{Yinan Hu}
\author[4]{Stefan Benders}
\author[5]{Mykhailo Flaks}
\author[5,6,7,8]{Dmitry Budker}
\author[2,*]{Alexej Jerschow}
\affil[1]{Department of Chemical and Biomolecular Engineering, NYU Tandon School of Engineering, Brooklyn, New York 11201, USA}
\affil[2]{Department of Chemistry, New York University, New York, New York 10003, USA}
\affil[3]{Institute of Biophysics, Chinese Academy of Sciences, Bei{}jing 100101, China}
\affil[4]{Institute of Process Imaging, Hamburg University of Technology, Hamburg, Germany}
\affil[5]{Institute of Physics, Johannes Gutenberg University Mainz, 55099 Mainz, Germany}
\affil[6]{Helmholtz Institute Mainz, 55099 Mainz, Germany}
\affil[7]{GSI Helmholtzzentrum f\"ur Schwerionenforschung, 64291 Darmstadt, Germany}
\affil[8]{Department of Physics, University of California, Berkeley, California 94720, USA}
\affil[*]{Corresponding author: alexej.jerschow@nyu.edu}
\date{25 November 2026}

\begin{document}
\maketitle

\begin{abstract}
Internal leakage currents are difficult to detect and to localize through measurements at the terminals of lithium-ion cells. We measured the magnetic fields of pouch cells non-invasively with a magnetometer while a battery cycler set the current and recorded the voltage. A three-axis magnetoresistive sensor remains fixed inside a three-layer magnetic shield, and two linear stages move the cell in two dimensions relative to the sensor to acquire a map of the magnetic field above the cell. The field of an intact cell was linear in current from 10 to 420 mA, and its maps were symmetric between charge and discharge. To create internal leakage currents, two 600 mAh cells were damaged by mechanical indentation. Both cells developed a magnetic field at zero applied current that decayed over hours, while the cell voltage showed a loss of charge by self-discharge. This field breaks the symmetry between the charge and discharge maps, which allows leakage to be detected while the cell is being cycled. At rest, the setup detects leakage currents down to about 1--3 mA. The results highlight opportunities for battery diagnostics and safety, and may help in designing techniques for characterizing soft shorts.
\end{abstract}

\section{Introduction}

Internal battery defects induced for example by mechanical deformation can damage the separator of a lithium-ion cell and
can cause an internal short circuit \cite{r4}, a known route to capacity loss and potentially thermal
Damage or improper construction can also leave a cell that discharges slowly but keeps operating
runaway \cite{r1,r2,r3}. Abuse testing standards test the response to harsh damage \cite{r5}.
\cite{r6,r7,r8,r9,r10,r11,r12,r17}. It is often more important to diagnose mild defects, such soft shorts with a weak
electrical signature \cite{r13,r14} which can change over time
\cite{r15,r16}.

Charge balance, float charging and potentiostatic holds measure leakage
currents of milliamperes or less \cite{r18,r19,r20}. These terminal
methods integrate over the whole cell and are complicated by voltage
relaxation \cite{r38,r21}, and they give no spatial information about a
leak, nor would they reflect on internal rebalancing currents that are not visible at terminals.

Magnetic measurements add this spatial information noninvasively
\cite{r22,r23,r24,r25,r26,r27}. Magnetic resonance imaging of the field
around a cell maps the susceptibility of the electrode materials, and
with it the state of charge and mechanical defects \cite{r22,r28}. It has
been extended to the measurement of current distributions \cite{r29},  to operating
commercial cells \cite{mohammadi2019persp,romanenko2020,romanenko2023}
Surface-scan variants reduce the hardware requirements
\cite{romanenko2022jpcc}. Direct magnetometry dispenses with the magnet.
Optically pumped magnetometers have mapped the induced magnetization of
cells, including cells with mechanical defects
\cite{r23,hu2020apsci}, and found weak transient fields at open circuit
\cite{r23,r35}, and nitrogen-vacancy sensors have imaged cells through
induced eddy currents \cite{zhang2021}. Field maps recorded under an
applied current have imaged current distributions \cite{r30,r31,r32} and
located manufacturing defects and simulated faults \cite{r25,r26,r32},
and a shielded scanning setup with an atomic magnetometer has mapped
fresh and over-discharged pouch cells of a similar model \cite{sharma}.
Leakage currents are harder to detect. The current of a local short
flows in nearly mirror-image patterns in neighboring collector foils,
whose fields largely cancel at the sensor (Sec.~III D), and
current-reconstruction methods have reached a detection limit of about
50 mA \cite{r27,r33}. Prior work has not followed, at both with and without applied current and
over time, the field of a mild internal leak  in a cell damaged to
mimic a mild short.

Here we describe a scanning magnetometer built around a magnetoresistive sensor and validate it with calibration measurements and an intact cell. We then indent cells in place and follow the field of the resulting internal leak at open circuit, without an applied magnetic field, starting minutes after the damage. We compare this field with the leakage current and charge loss determined from the cell voltage, and calibrate the sensitivity to leakage current at 1--3 mA. Unlike Ref.~\citenum{r23}, which measured induced magnetization in an applied field and transient currents in intact cells, we follow the field that appears after damage both with and without applied current and compare measurements with the leakage current and the charge loss recorded by the cycler.

\section{Experimental setup}

\subsection{Scanning magnetometer}\label{sec:setup}

\begin{figure}[tb]\centering\includegraphics[width=\textwidth]{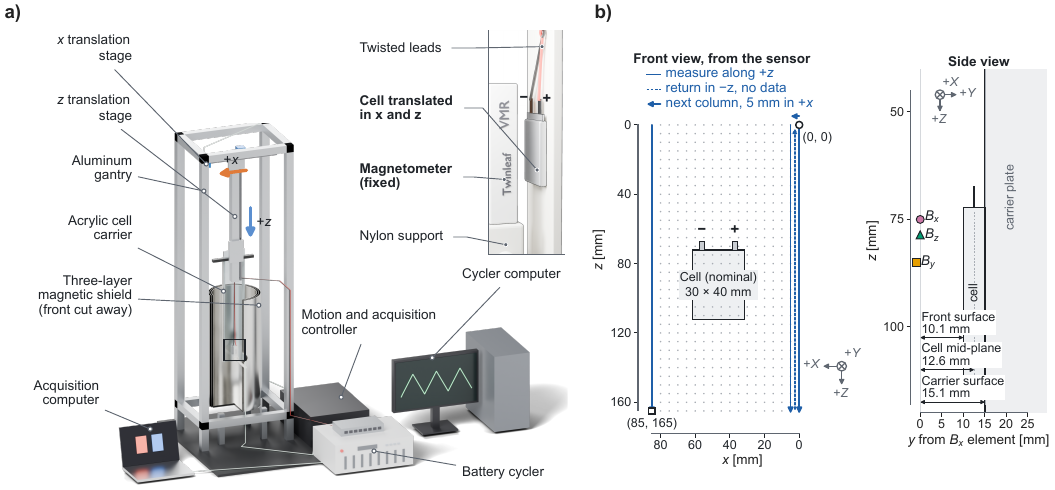}\caption{Scanning magnetometry setup. (a) Rendering with approximate proportions. Two linear stages outside a three-layer magnetic shield move the cell platform in the $x$ and $z$ directions past a fixed three-axis magnetometer (inset). The front of the shield is drawn cut away to show the cell and the magnetometer inside. A battery cycler sets the current and records the voltage. (b) Scan grid (18 $\times$ 34 points, 5 mm spacing) and cell position seen from the sensor, registered with the wire measurements. Side view: distances from the $B_x$ sensing element to the cell, from field fits.}\label{fig:1}\end{figure}

Figure~\ref{fig:1}(a) shows the magnetometry setup. A three-axis magnetoresistive
magnetometer sensor (Twinleaf VMR; 300 pT Hz\(^{-1/2}\), \(\pm\)100
\textmu T range) is attached on a nylon support inside a three-layer
cylindrical mu-metal shield (Twinleaf MS-2) with bottom caps. The top cap was removed to allow the cell platform to shuttle through the top opening. Two layers of permalloy tape (grade 1J85, about 80~\% Ni) cover the opening, apart from the clearance the platform needs to pass. Covering the opening reduced the scan-to-scan noise in $B_z$ from 14.9 nT (opening uncovered) to 4.4 nT.

The cell is fixed to an acrylic platform that two orthogonal linear stepper stages shuttle along \(x\) and \(z\), in the plane facing the sensor. This gives a two-dimensional map of the magnetic field over the cell. A scan covers an 18-by-34-point grid with 5 mm spacing [Fig.~\ref{fig:1}(b)]. Before each reading, the stages pause for 0.2 s so that vibrations of the long carrier die down, and two consecutive magnetometer readings, taken at an output rate of 10 Hz, are averaged. One scan of the 85 mm \(\times\) 165 mm grid takes about 9 min. The first scan after homing is shifted by 0.3--0.7 mm in \(x\), due to imperfections in the homing procedure in some scans, and we discard this scan where it appears. \(B_y\) is normal to the cell face, and \(B_x\) and \(B_z\) lie in the scan plane. Repeat scans without current differ by 1.1, 0.7 and 4.2--4.4 nT (standard deviation) in \(B_x\), \(B_y\) and \(B_z\) [Fig.~S1]. Table~\ref{tab:noise} lists this and the other noise figures used below, each for its own quantity.

Without a sample, the readings of the magnetic field strength (residual field plus sensor offsets) were
515, \(-27\) and 15 nT in \(B_x\), \(B_y\) and \(B_z\). Across the
grid, the mean of three such scans varies by 0.7, 0.4 and 1.9 nT
(standard deviation), no more than the scan noise, so the residual field
is uniform over the scan area and cancels in background-subtracted maps. The residual field is at most about 0.5~\textmu T, so a change in cell susceptibility of order \(10^{-4}\) (SI) changes the measured field by less than 0.1 nT.
Maps of cell D1 without current before indentation and 30 and 77 h after it agree at the noise level (Sec.~\ref{sec:d1}). The 5 mm grid and the 10 mm
distance to the cell limit the spatial resolution to features of about
10 mm. At the intact-cell response of 10.23 nT mA\(^{-1}\)
(Sec.~\ref{sec:validation}), the 0.7 nT rms noise of \(B_y\) corresponds
to 0.07 mA of terminal current.

The scan is run from a Jupyter Python notebook on a Red Pitaya board, which also reads the magnetometer over USB through Twinleaf's background data server and Python package (tldevice). The board sends step and direction pulses to TN600 stepper-motor drivers (24 V supply). One step moves a stage by 0.02 mm. Before each scan, both stages are driven to the top left limiting position (determined by limit switches) to set the zero position (homing). Each column is measured point by point along $z$, and the stage returns without measuring before stepping 5 mm in $x$ [scan path in Fig.~\ref{fig:1}(b)]. One measurement point takes about 0.9 s. The battery cycler runs on a separate computer, and the scans and cycler records are matched by their time stamps.

\begin{table}[tb]\centering\small\renewcommand{\arraystretch}{1.12}
\caption{Noise figures used in this work. All are measured without applied current.}\label{tab:noise}
\begin{tabular}{L{0.47\textwidth}L{0.17\textwidth}L{0.27\textwidth}}
\hline\hline
Quantity & Value & Use \\
\hline
Standard deviation of each grid point between repeat scans (\(B_x\), \(B_y\), \(B_z\)) & 1.1, 0.7, 4.2--4.4 nT & scan noise [Fig.~S1] \\
Standard deviation over the grid of the mean of three scans without a sample (\(B_x\), \(B_y\), \(B_z\)) & 0.7, 0.4, 1.9 nT & uniformity of the residual field \\
Terminal current equivalent of the 0.7 nT \(B_y\) noise at 10.23 nT mA\(^{-1}\) & 0.07 mA & sensitivity to applied current \\
Rms over the map of D1 at 30 and 77 h minus the fresh cell (\(B_y\)) & 0.6 nT & return to the fresh state (noise alone: 0.6 nT) \\
Scan-to-scan rms of the fitted D2 amplitude & 0.28 nT & amplitude noise \\
Detection limit, three times the amplitude noise & 0.85 nT & sensitivity to leakage current \\
\hline\hline
\end{tabular}
\end{table}

\subsection{Calibration}

Two copper wires were laid on the carrier at the position of the cell, one along \(z\) and one along \(x\), with leads running vertically to the top and bottom of the scan area. A current of 60 mA, measured with a series ammeter (Fluke 115), was passed through each wire. The fit to the wire along \(z\) (Sec.~\ref{sec:validation}) places the wire center 14.1 mm from the \(B_x\) sensing element. With the 1 mm wire radius, the carrier surface is 15.1 mm from the element. The cell is 5 mm thick, so its mid-plane is at 12.6 mm and its front surface at 10.1 mm [Fig.~\ref{fig:1}(b)]. The wire positions also locate the cell outline on the grid.

The three field components are measured by separate sensing elements at different positions along \(z\). Between the cell and the sensor there are no currents or magnetic materials, so the field measured is given by  the gradient of a scalar potential, and the map of one component, \(B_y\), in principle determines the maps of the other two up to a uniform offset. Fitting this relation to the maps places the \(B_y\) and \(B_z\) elements \(10.1\pm0.2\) and \(3.6\pm0.3\) mm from the \(B_x\) element along \(z\), and less than 1 mm from it along \(x\). The same fit fixes the sign of \(B_z\). Because the maps are recorded in 5 mm steps, the \(B_y\) and \(B_z\) maps were shifted along \(z\) by the nearest whole steps, 10 and 5 mm. The remaining offsets, 0.1 and 1.4 mm, are below a third of a step. Every amplitude in this work is fitted with a pattern taken from maps registered in the same way, so these offsets do not change the amplitudes. The top rows, which have no data after this shift, are not shown in the maps.

\subsection{Cells and indentation}

Three PGEB-NM053040 pouch cells (600 mAh, 2.22 Wh, 30 mm × 40 mm) were used: one intact and two indented (D1 and D2, D for dented). All three were new and at 3.78 V when mounted. With the cell taped to the platform, a 9 mm diameter rod was struck with a hammer through a plastic sheet and paper towels. This left a circular dent about 9 mm across near the center of the cell face, approximately 3 mm deep. All measurements were made at room temperature, which was not logged.

A Neware
cycler with four-wire connections logged current and voltage every
second. The cycler recorded the actual current as 179.8--179.9 mA in the log file (nominal current setting was 180 mA). 

\subsection{Protocols}

\emph{Intact cell:} 33 sessions, each a single charge or discharge at constant current, were run over seven days at 10--480 mA. Each session had one to four scans, depending on the current. A scan of the mounted cell without current gave the background field map. Twenty-eight sessions had at least one complete scan after the first, with the current confirmed by the cycler record. The other five were not fitted. The 420 mA charge and the 480 mA charge and discharge have only a first scan, which is not used. The 60 mA charge has no cycler record, and no scan of the 60 mA discharge on 20 October was fully under current.

\emph{Cell D1:} Scans without current, then three scans each during a
180 mA charge and discharge. After indentation, three scans each during
a 180 mA discharge and charge from 0.2 h, then three scans without
current at 1.7--2.0 h. The cycler was then off for 1.7 h, followed by
five cycles between 3.0 and 4.2 V at 180 mA [Fig.~\ref{fig:5}(a)]. Scans
without current were repeated at 30 h and, after a second run of 180 mA cycles (17 cycles in total) [Fig.~S3], at 77 h.

\emph{Cell D2:} One rest step at 0 mA for 18.7 h, with four scans before
indentation and 100 scans from 0.1 to 15.5 h after it.

\subsection{Separating the field of the applied current}\label{sec:separation}

The field at each point is
\begin{equation}
\mathbf{B} = I\,\mathbf{b} + \mathbf{B}_0, \label{eq:decomp}
\end{equation}
where \(\mathbf{b}\) is the field per unit charging current and
\(\mathbf{B}_0\) is the current-independent field. \(\mathbf{B}_0\) contains the background, noise, magnetization and the fields of internal currents. Combine charge and discharge maps
as
\begin{equation}
\begin{aligned}
\tfrac{1}{2}\left[\mathbf{B}(+I)-\mathbf{B}(-I)\right] &= I\,\mathbf{b} + \tfrac{1}{2}\,\Delta\mathbf{B}_0,\\
\tfrac{1}{2}\left[\mathbf{B}(+I)+\mathbf{B}(-I)\right] &= \overline{\mathbf{B}}_0,
\end{aligned}\label{eq:oddeven}
\end{equation}
where \(\Delta\mathbf{B}_0\) is the change of \(\mathbf{B}_0\) between the two maps and \(\overline{\mathbf{B}}_0\) is its mean. The half difference therefore gives the field of the applied current, and the half sum gives the current-independent field. Scans without applied current measure \(\mathbf{B}_0\) directly and were compared with the half sum. Each map was background-subtracted and centered over a mask that
excludes the outer two to three rows and columns. No smoothing was applied. For D1 scans under current, the fresh applied-current field
(free amplitude and sub-pixel shift) and the pattern of the scans without
current at 1.7--2.0 h were fitted together. D2 amplitudes were fitted to
the mean of its first three post-indentation maps, using \(B_x\) and
\(B_y\). Amplitudes are the \(B_y\) peak of the fitted pattern. We determined the
detection limit as 0.85 nT, which is three times the scan-to-scan rms noise of this
amplitude (0.28 nT).

\begin{figure}[tb]\centering\includegraphics[width=\textwidth]{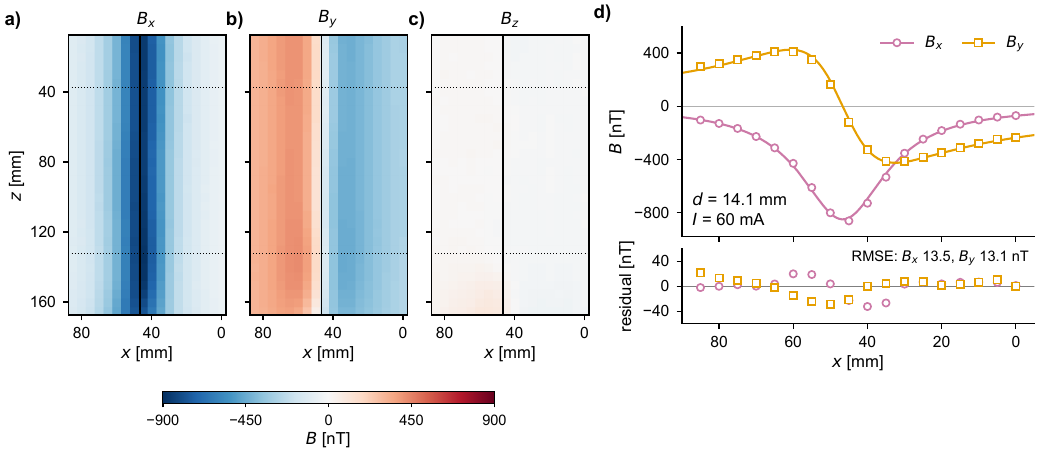}\caption{Wire along $z$ carrying 60 mA. (a)--(c) $B_x$, $B_y$ and $B_z$ maps after subtracting a map without current. Solid line: fitted wire position; dotted lines: rows averaged in (d). (d) Measured $B_x$ and $B_y$ profiles (symbols), straight-wire fit with the current fixed at the measured 60 mA and free distance (lines), and residuals.}\label{fig:2}\end{figure}

\subsection{Leakage current from the cell voltage}\label{sec:voltage}

To compare the magnetic field with the leakage current, we estimate the current independently from the cell voltage. A cell with an internal leak loses charge even at rest, so its voltage falls. The leakage current follows from the rate of this fall and the charge stored per volt:
\begin{equation}
I_\mathrm{L} = -\,\frac{\mathrm{d}Q}{\mathrm{d}V}\,\frac{\mathrm{d}V}{\mathrm{d}t}. \label{eq:ileak}
\end{equation}
We use this for D2, whose voltage was recorded every second at rest. The rate \(\mathrm{d}V/\mathrm{d}t\) is taken from linear fits to the voltage in consecutive windows (10 min up to 2 h, 30 min up to 6 h, 1 h thereafter) and, for comparison with the field, over each scan. The differential capacity \(\mathrm{d}Q/\mathrm{d}V\) is taken from D1's fifth charge--discharge cycle, since both cells are of the same model. During cycling, the applied current raises the voltage while charging and lowers it while discharging, and a cell at rest lies between the two curves. We therefore use the mean of the two curves, fitted with a fourth-order polynomial between 3.45 and 3.85 V. Over D2's voltage range this gives 675--906 mAh V\(^{-1}\). Using only the discharge curve gives currents 0.58--0.81 times as large. The charge curve does not reach below 3.66 V, where most scans lie. This shows how sensitive the estimate is to this choice. With the discharge curve, the field per milliampere of leakage in Sec.~\ref{sec:d2} would be 1.2--1.7 times larger. D1 had already lost capacity by its fifth cycle (338 mAh discharged, against 600 mAh nominal), so if D2 kept more capacity, its leakage currents would be correspondingly larger.

For D1, the voltage was not recorded during its rests, so we use a second estimate that needs no \(\mathrm{d}Q/\mathrm{d}V\). The rest voltage of a cell rises with the charge it stores. A cell that receives a charge \(\Delta Q\) and afterwards rests at a lower voltage than before must therefore have lost at least \(\Delta Q\). This holds as long as the relation between rest voltage and stored charge does not change and no current is drawn externally. The two estimates have different strengths. The first follows the leakage current in time but depends on the choice of \(\mathrm{d}Q/\mathrm{d}V\). The second gives only a lower bound on the mean current over the rest, but needs no such choice. They could not be compared on one cell, because the voltage of D1 was not recorded at rest and D2 received no charge.

\begin{figure}[tb]\centering\includegraphics[width=0.90\textwidth]{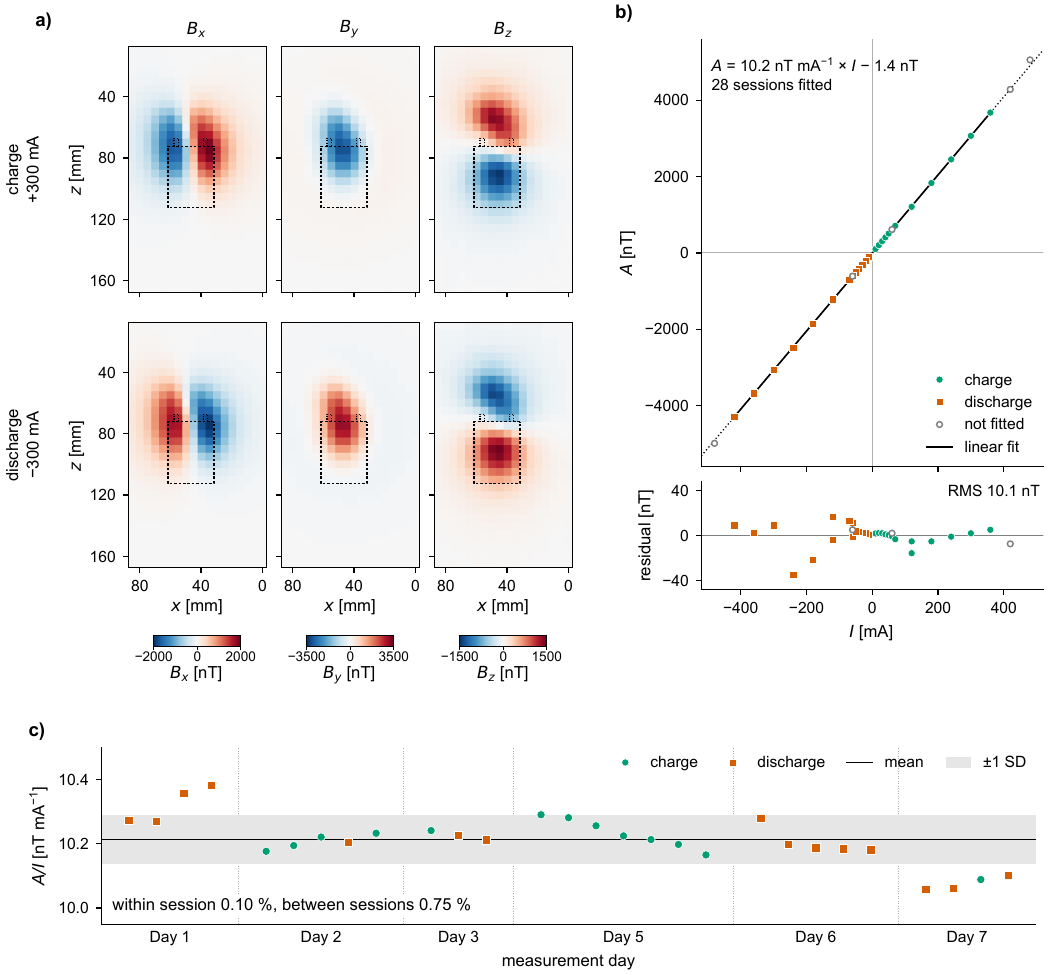}\caption{Intact cell. (a) $B_x$, $B_y$ and $B_z$ maps at $+300$ mA (charge, top) and $-300$ mA (discharge, bottom); dashed line: cell outline. (b) Amplitude $A$, the size of the $B_y$ field pattern, of each charge and discharge map against current $I$, with a linear fit; open symbols: not fitted. (c) $A/I$ of each charge and discharge map by day; line: mean; band: $\pm$1 standard deviation.}\label{fig:3}\end{figure}

\section{Results and discussion}

\subsection{Validation}\label{sec:validation}

The wire along \(z\) calibrates the distance between platform and field sensing location, and the wire along \(x\) is further used for verification. With the current fixed at the measured 60 mA, a straight-wire fit to the wire along \(z\) gives a distance of 14.1 mm, with rms residuals of 1.6 and 3.2~\% of the peak \(B_x\) and \(B_y\) profiles [Fig.~\ref{fig:2}]. \(B_z\), the component along the wire, stays below 96 nT, against 888 nT peak \(B_x\). With 60 mA and 14.1 mm fixed, the wire along \(x\) reproduces \(B_z\) with an rms residual of 1.3~\% [Fig.~S2].

\begin{figure}[tb]\centering\includegraphics[width=0.90\textwidth]{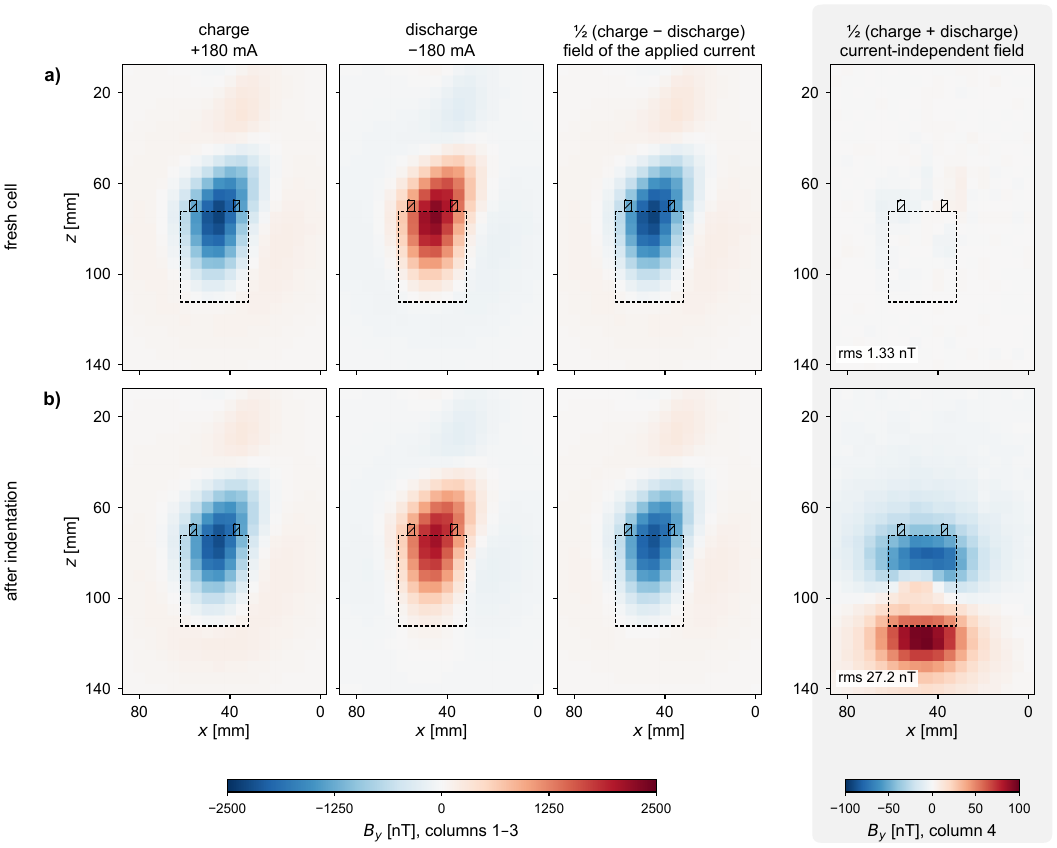}\caption{Separation of the applied-current and current-independent fields ($B_y$, cell D1). Rows: (a) fresh cell; (b) after indentation. Columns: charge ($+180$ mA), discharge ($-180$ mA), half difference (field of the applied current) and half sum (current-independent field). Color scales: $\pm$2500 nT (columns 1--3) and $\pm$100 nT (column 4).}\label{fig:4}\end{figure}

Figure~\ref{fig:3}(a) shows the intact cell at $\pm$300 mA. Charge and discharge give the same field pattern with opposite sign: fitting the discharge map to the charge map point by point gives a slope of $-0.997$ for \(B_y\).

To compare charge and discharge maps, we describe each \(B_y\) map by one number, its amplitude \(A\). We fit the map with a fixed reference pattern \(T\), the average shape of the other sessions scaled to a peak of 1:
\begin{equation}
B_y(x,z) = A\,T(x,z) + c,
\end{equation}
where \(c\) is a constant offset. Because the shape stays the same, \(A\) gives the size of the field, positive for charge and negative for discharge.

From 10 to 420 mA, the amplitude grows in proportion to the current, \(A = kI + A_0\), with \(k = 10.23\) nT mA\(^{-1}\) and \(A_0 = -1.4\) nT [Fig.~\ref{fig:3}(b)]. The five sessions shown as open symbols were not fitted (Sec.~II D). The field per unit current, \(A/I\), varies by 0.75~\% (standard deviation between sessions) over six measurement days [Fig.~\ref{fig:3}(c)]. The sensor is less sensitive than the atomic magnetometer of Ref.~\citenum{sharma} (15 fT Hz\(^{-1/2}\)), but its \(\pm\)100~\textmu T range covers the fields of the currents used to cycle the cells.

\begin{figure}[tb]\centering\includegraphics[width=0.90\textwidth]{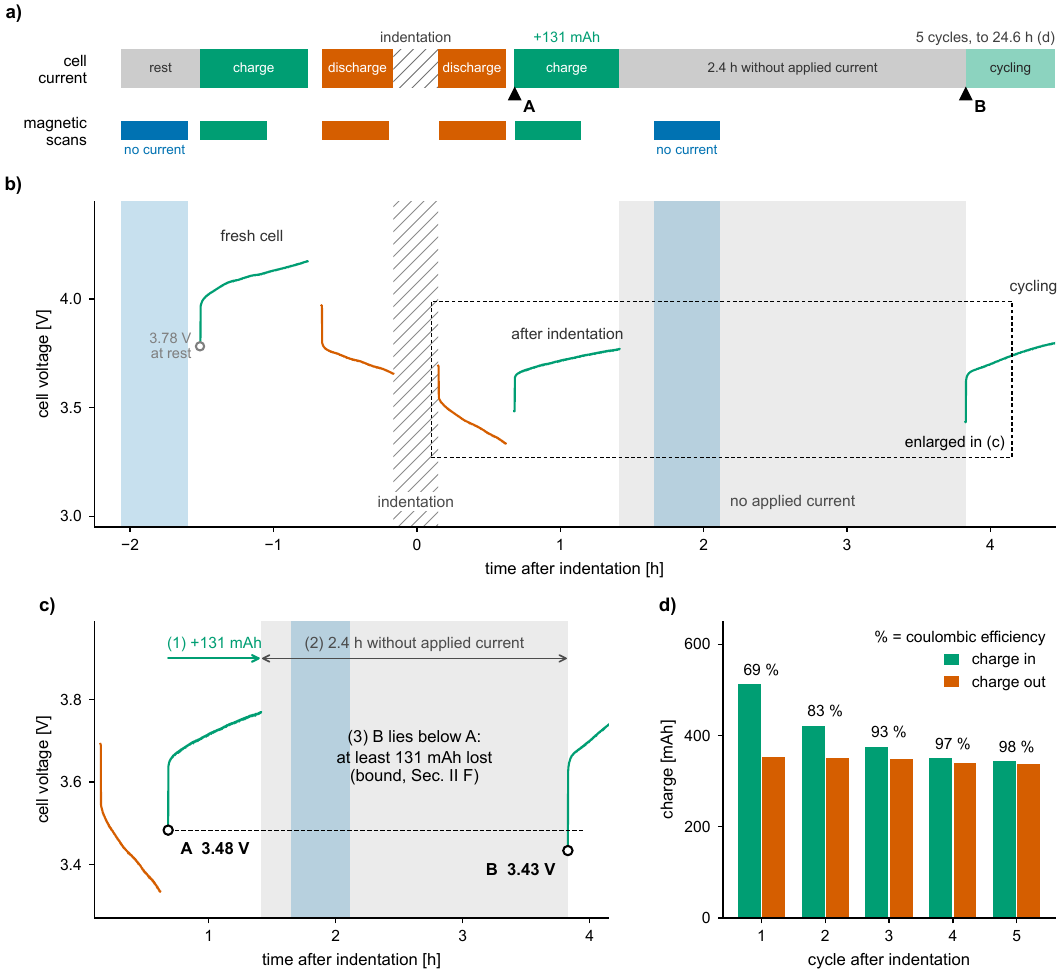}\caption{Cell D1, electrical record. (a) Cycler steps on the day of indentation (top) and blocks of three scans (bottom). (b) Cell voltage on the same time axis; open circle: initial rest voltage (3.782 V); dashed box: region enlarged in (c). No voltage was recorded during the scans without current. (c) Rest voltage before a 131 mAh charge (A, 3.483 V) and after it plus 2.4 h without current (B, 3.434 V). (d) Charge in and out, with coulombic efficiency, for the five cycles after B.}\label{fig:5}\end{figure}

\subsection{Cell D1: indentation between current blocks}\label{sec:d1}

Figure~\ref{fig:4} compares the maps of D1 before and after indentation. Before indentation, the charge and discharge maps are mirror images. Their half difference is the field of the applied current, with a peak of 2.3~\textmu T and an rms of 512 nT over the map, and their half sum is close to zero (1.3 nT rms). After indentation, the field of the applied current keeps its form (2.2~\textmu T peak, 478 nT rms). The half sum, however, rises to 27 nT rms, with a peak near 100 nT, and shows a two-lobed pattern that was absent before. The fresh half sum, 1.3 nT rms with a 5.6 nT peak, includes what positioning errors leave and is far below the 100 nT peak after indentation. Charge and discharge maps are therefore no longer mirror images, and this loss of symmetry is the first indication of the leak. The half sum is useful because it can be formed while the cell is being cycled. At rest, a scan without current gives the same field more directly, without cancelling the 2.3~\textmu T field of the applied current.

\begin{figure}[tb]\centering\includegraphics[width=\textwidth]{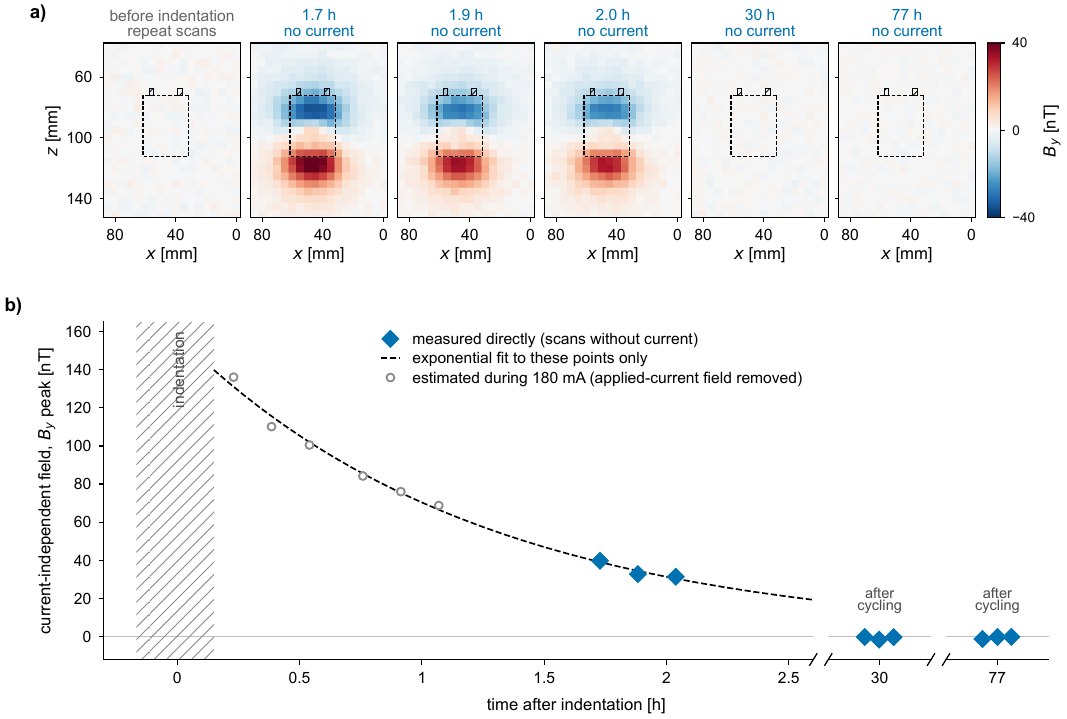}\caption{Cell D1, current-independent field. (a) $B_y$ maps of scans without current, minus the fresh-cell scans: before indentation (difference of two repeat scans, showing the noise) and 1.7, 1.9, 2.0, 30 and 77 h after it. (b) $B_y$ peak amplitude against time; the exponential fit (dashed) uses only the scans without current. Estimates from the 180 mA scans, after removal of the background and of the refitted applied-current field, were not fitted and lie within 5~\% of the fit's extension. The 30 h and 77 h scans were performed after cycling.}\label{fig:6}\end{figure}

The cell voltage confirms that D1 was leaking [Fig.~\ref{fig:5}]. Before a 131 mAh charge, D1 rested at 3.483 V. After the 0.73 h charge and 2.4 h of rest, it was at 3.434 V, lower than before charging [points A and B in Fig.~\ref{fig:5}(c)]. A cell that gains charge should rest at a higher voltage, so D1 must have lost at least the 131 mAh it received. Point A was taken just after a discharge and B after a charge and 2.4 h of rest. Relaxation raises the voltage after a discharge and lowers it after a charge, so it makes this bound conservative. Over these 3.1 h, this corresponds to a mean leakage current of at least 42 mA. The leakage then faded. Coulombic efficiency, the charge recovered on discharge divided by the charge put in, would be close to 100~\% without a leak, because charge lost inside the cell cannot be recovered. It rose from 69~\% in the first cycle to 98~\% in the fifth, as the charge lost per cycle fell from 159 to 6 mAh [Fig.~\ref{fig:5}(d)]. The leakage thus faded within the first day after indentation, as did the current-independent field, which had disappeared by 30 h [Fig.~\ref{fig:6}].

The current-independent field decayed over hours [Fig.~\ref{fig:6}].
Scans without current gave 31--40 nT at 1.7--2.0 h, and an exponential fit to these scans gives a time constant of 1.2--1.3 h. Values estimated during
the earlier scans under current, from 136 nT at 0.2 h to 69 nT at 1.1 h,
lie on the extension of this fit. Fits that include them give 1.1--1.3 h. At 30 and 77 h, maps without current (mean of two scans) differed from the fresh cell by 0.6 nT rms, the value expected from scan noise alone for this comparison [Fig.~\ref{fig:6}(a)]. The field was therefore not caused by a
permanent change, such as material left by the indenter.

\subsection{Cell D2: indentation at rest}\label{sec:d2}

\begin{figure}[tb]\centering\includegraphics[width=0.92\textwidth]{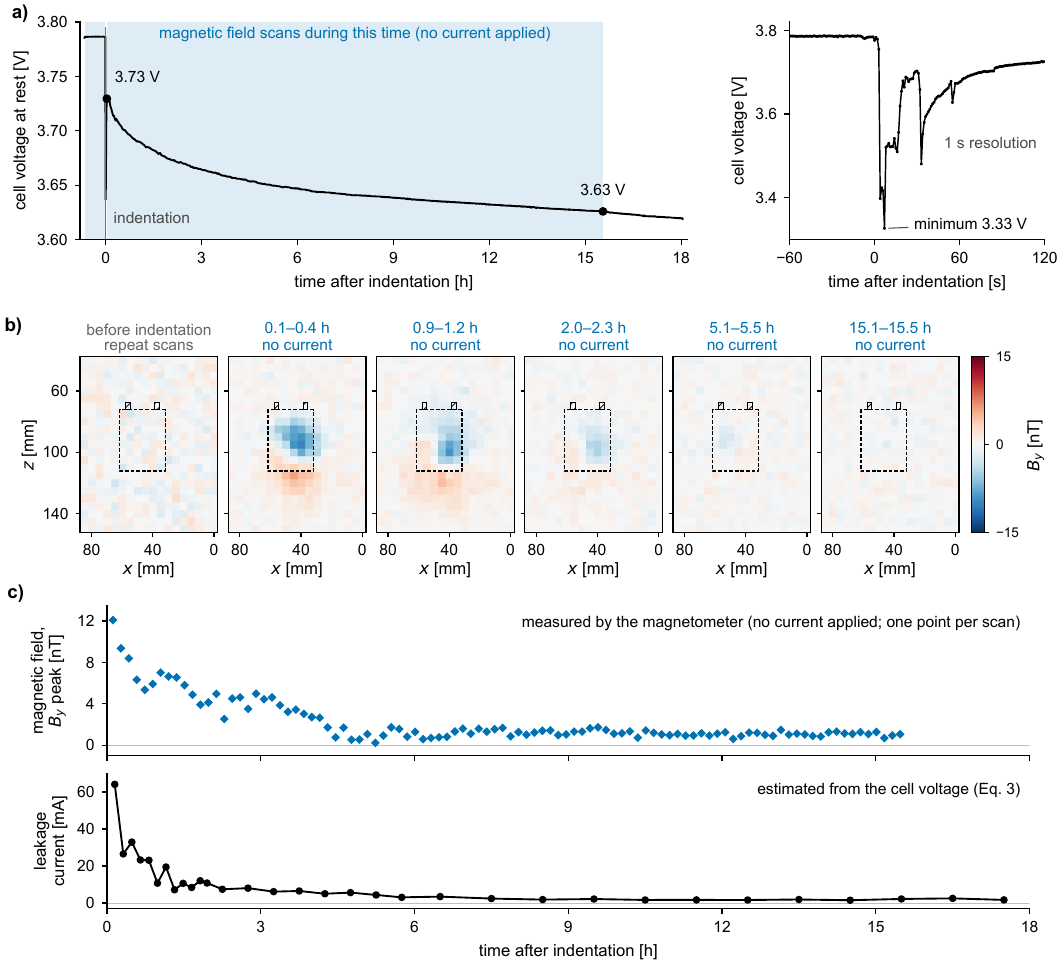}\caption{Cell D2, indented at rest with no applied current. (a) Cell voltage over 18.7 h (left) and around indentation at 1 s resolution (right); shaded band: the 100 scans during which the voltage fell from 3.73 to 3.63 V. (b) $B_y$ maps before indentation (difference of two repeat scans, showing the noise) and at five times after it (mean of three scans minus the pre-indentation mean). (c) $B_y$ peak magnitude in every scan (top) and leakage current from Eq.~(3) (bottom).}\label{fig:7}\end{figure}

D2 was not subjected to charging or discharging during measurements in order to get a more complete measurement set after indentation [Fig.~\ref{fig:7}]. Its voltage fell
from 3.7864 V before indentation to 3.728 V at 4 min and 3.619 V at 18 h.
Brief dips to 3.327 V in the first 55 s were excluded.
Equation~\eqref{eq:ileak} gives a leakage current of about 64 mA at
0.1 h, 12 mA at 1 h, 3 mA at 6 h and 2 mA at 14--17.5 h. The charge
lost is 122 mAh, of which 33 mAh fall in the least reliable first four
minutes, when the voltage was still recovering from the dips. At 64 mA and 3.7 V, about 0.24 W was dissipated at the short at 0.1 h.

The first map after indentation shows a two-lobed pattern with a \(B_y\)
peak of \(-12\) nT, absent before indentation. It declines to \(-8\) nT
over the first hour, \(-4\) nT at 2--3 h and about \(-1\) nT after 5 h. The fitted amplitude keeps its sign and was close to zero before indentation (\(-0.19\) to 0.26 nT in four scans). After 5 h, all 68 scans gave negative amplitudes with a mean of \(-1.1\) nT, so this residual is not scan noise, although slow drifts over the night cannot be fully excluded. It was not followed beyond 15.5 h.
The field amplitude follows the leakage current (\(r=0.87\); 0.26 nT
mA\(^{-1}\), offset 1.0 nT) and the voltage decline rate (\(r=0.86\)).
The first hours dominate this relation. Without the three scans that
define the fitted pattern, both correlations are 0.78. For all scans, an eight-scan
block bootstrap gives 95~\% intervals of 0.12--0.90 and 0.14--0.90, and
rank correlations are 0.48 and 0.49. An exponential decay to a constant offset (\(-1.0\) nT) gives 2.0 h for
the field (3.5 h without the offset), and the field and the leakage current decay with different
shapes. After 5 h, the field was close to the
0.85 nT detection limit while the voltage still indicated a few
milliamperes of leakage.

\subsection{Sensitivity and limitations}

Table~\ref{tab:1} compares the two cells. Between 0.7 and 3.8 h, D1 lost at least 131 mAh, a mean leakage current of at least 42 mA. Over the same period its field, taken from the exponential fit, averaged 34 nT, so each milliampere of leakage produced at most 0.81 nT. D2's field averaged 4.8~nT at 10~mA over the same period, or 0.47 nT mA\(^{-1}\), and 0.26--0.72 nT mA\(^{-1}\) over all its scans. This is 16--50 times less than the field that fresh D1 produced per milliampere of applied current, 12.9 nT mA\(^{-1}\). Both the applied and the leakage current run in opposite directions in neighboring foils \cite{r27,r33}, so the difference is most likely geometric. The tabs are separated in the plane, so the applied current leaves an uncancelled current loop about the size of the tab spacing. A local short instead draws current radially in one foil and returns it radially in the other. These nearly mirror-image patterns produce little field, and what is measured comes from what breaks the symmetry, such as the cell edges, an off-center short and the current that the rest of the cell feeds to the short through the collectors. The 12.9 nT mA\(^{-1}\) of D1 is 26~\% above the intact cell's 10.23 nT mA\(^{-1}\). The two cells are of the same model but were mounted separately, weeks apart, so a difference in stand-off or between the cells could explain it. We did not separate the two, and therefore compare the leakage fields of D1 and D2 only with D1's own value, measured in the same campaign. With the 0.85 nT detection limit, the setup can detect leakage currents at rest down to about 1--3 mA.

\begin{table}[tb]\centering\small\renewcommand{\arraystretch}{1.12}
\caption{The two indented cells. Electrical values are from the cell voltage (Sec.~\ref{sec:voltage}).}\label{tab:1}
\begin{tabular}{L{0.34\textwidth}L{0.27\textwidth}L{0.27\textwidth}}
\hline\hline
 & D1 & D2 \\
\hline
Applied current & \(\pm\)180 mA & none \\
Charge loss & \(\geq\)131 mAh in 3.1 h & 122 mAh in 18 h \\
Leakage current & \(\geq\)42 mA (0.7--3.8 h) & 64 mA to 2 mA \\
Current-independent field (\(B_y\) peak) & 31--40 nT at 1.7--2.0 h & \(-12\) nT at 0.1 h \\
Decay time & 1.1--1.3 h & 2.0 h (exponential plus offset) \\
Field per leakage current & \(\leq\)0.81 nT mA\(^{-1}\) & 0.26--0.72 nT mA\(^{-1}\) \\
\hline\hline
\end{tabular}
\end{table}

In D2, the magnetic field followed the leakage current estimated from the cell voltage. Charge redistribution after local discharge could also contribute to the current-independent field \cite{r22,r23}, while magnetization of the cell in the residual field adds less than 0.1 nT (Sec.~\ref{sec:setup}). Redistribution could also be set off by the damage and fade with time, so it cannot be separated from the field of the leak here. The voltage of D2 dipped briefly, to as low as 3.327 V, in the first 55 s, which suggests short-lived hard shorts. Each would discharge the region near the dent, and the resulting gradient in state of charge would relax through internal currents over hours even without a sustained leak, as seen in intact cells after charging \cite{r23}. If present, redistribution would change the field per milliampere of leakage in Table~\ref{tab:1}, but not the main observation: the field appeared with the leak and faded as the leak weakened. Scanning an intact cell at rest after the same 180 mA history as D1 would test how much of the field comes from such redistribution. The leakage current itself comes from the cell voltage, not from an independent measurement, and a fading field shows that the leak weakened, not that the cell recovered \cite{r39}.
\section{Conclusion}

A magnetoresistive scanning magnetometer is described that allows detecting internal leakage currents of lithium-ion pouch cells through their magnetic fields. Magnetic field patterns in intact cells were shown to be linear with applied current and exhibited a symmetry between charge and discharge operation. After indentation,  cells developed a current-independent field that
decayed over hours while the voltage showed charge loss, and broke the symmetry between charge and discharge measurements. In the measurement setup described here, the detection limit of leakage current was determined to be
about 1--3 mA. With a sensor as sensitive as the atomic magnetometer of Ref.~\citenum{sharma} (15 fT Hz\(^{-1/2}\), against 300 pT Hz\(^{-1/2}\) here), sensor noise alone would allow leakage currents in the sub-\textmu A range. Lower currents would also need better scan reproducibility and positioning, and an understanding of the redistribution background. A longer dwell per point should already lower the noise of the present sensor. The scan time was not a limiting factor in this one-sensor setup, but further speed-up could be achieved with sensor arrays. With fewer stage movements, less time would also be spent waiting for vibrations to settle. These types of measurements could become important techniques for battery diagnostics, especially for detecting soft shorts. 

\section*{Supplementary Material}

See the supplementary material for the scan noise [Fig.~S1], the validation with the wire along \(x\) [Fig.~S2], the cycling of cell D1 after indentation [Fig.~S3] and photographs of the indented cells D1 and D2 [Figs.~S4 and S5].

\section*{Acknowledgments}
A.J. acknowledges funding from the US National Science Foundation (Award No. CHE 2505792), an award from the ACS (PRF 68117-ND6), and the Carl Zeiss Humboldt Research Award. This study was supported in part by the Helmholtz Initiative and Networking Fund (project KA-QUS-03 -- QT-Batt).

\section*{Author Declarations}
\subsection*{Conflict of Interest}
The authors have no conflicts to disclose.

\section*{Data Availability}

Data supporting the findings of this study are available from the corresponding author upon reasonable request.

\renewcommand{\refname}{REFERENCES}
{\small\setlength{\itemsep}{1pt}
}

\end{document}


\begin{center}
{\Large\bfseries Supplementary Material}\\[0.5em]
{\large Diagnosing internal leakage currents through magnetometry of lithium-ion cells}\\[0.5em]
Shaurya Shukla, Chengtong Zhang, Yinan Hu, Stefan Benders, Mykhailo Flaks, Dmitry Budker, Alexej Jerschow
\end{center}

\begin{figure}[H]\centering\includegraphics[width=\linewidth,height=0.33\textheight,keepaspectratio]{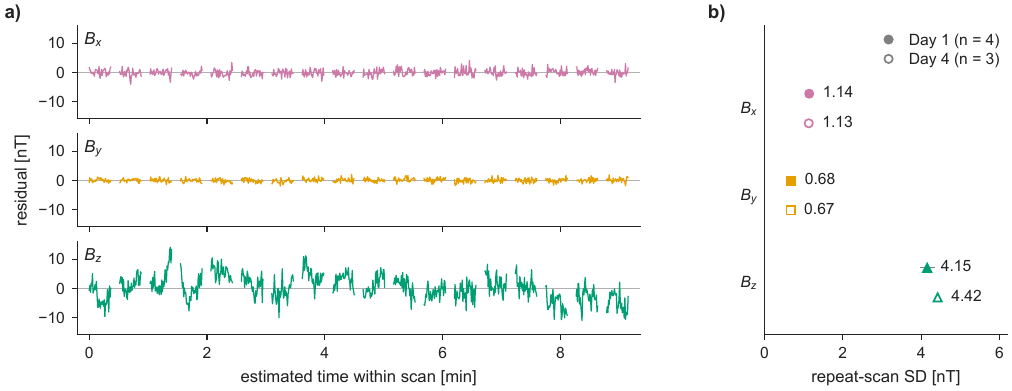}\caption{Scan noise without current. (a) Deviation of one scan from the daily mean against time within the scan. (b) Standard deviation between repeat scans on two days, 14 days apart ($n = 4$ and 3).}\label{fig:S1}\end{figure}

\begin{figure}[H]\centering\includegraphics[width=\linewidth,height=0.33\textheight,keepaspectratio]{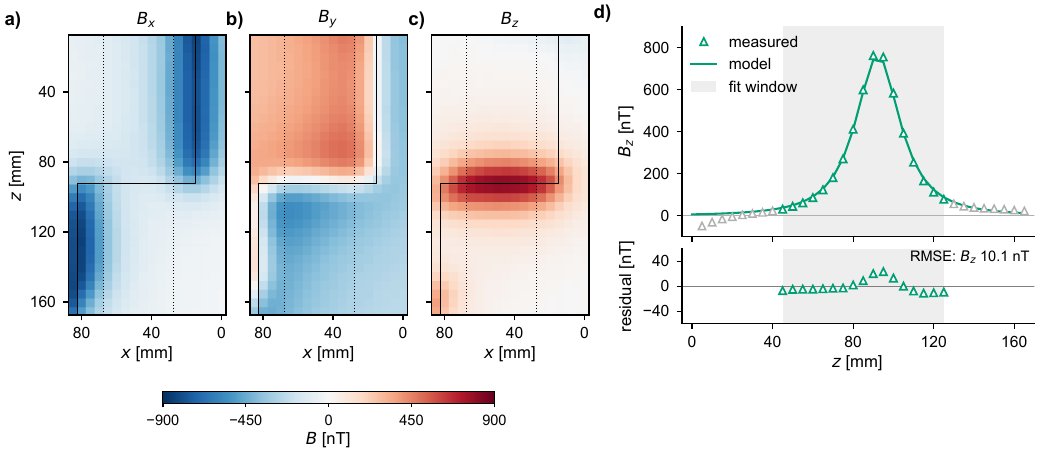}\caption{Wire along $x$ carrying 60 mA. (a)--(c) Maps after subtracting a map without current; solid line: wire path; dotted lines: columns averaged in (d). (d) Measured $B_z$ profile and finite-length wire field (current fixed at 60 mA and distance fixed at 14.1 mm from Fig.~2 of the main text), with residuals; shaded: fit window. Only the wire alignment and column offsets were fitted (residual 10.1 nT).}\label{fig:S2}\end{figure}

\begin{figure}[H]\centering\includegraphics[width=\linewidth,height=0.30\textheight,keepaspectratio]{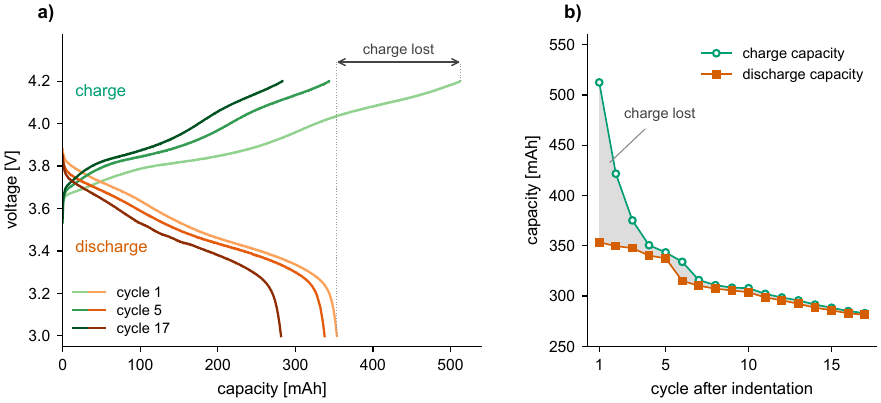}\caption{Cell D1 cycled after indentation. (a) Voltage versus capacity over different cycles. In cycle 1, much more charge went in than came out; by cycle 5 the two nearly match. The curve shape stays the same, but the capacity drops with each cycle. (b) Charge and discharge capacity of each cycle; shaded: charge lost.}\label{fig:S3}\end{figure}

\begin{figure}[H]\centering
\begin{minipage}[t]{0.535\linewidth}\centering\includegraphics[width=\linewidth]{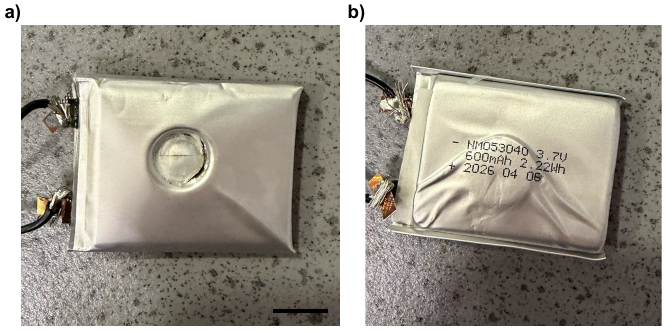}\caption{Cell D1 after indentation. (a) Front face; scale bar, about 10 mm. (b) Back face.}\label{fig:S4}\end{minipage}\hfill
\begin{minipage}[t]{0.445\linewidth}\centering\includegraphics[width=\linewidth]{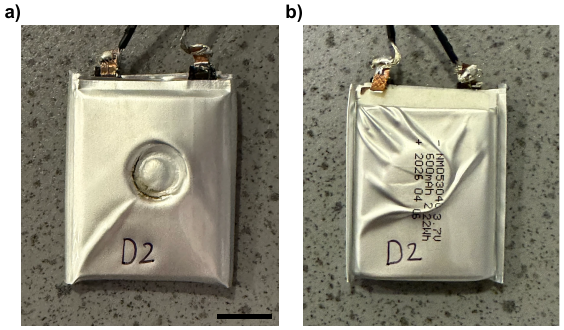}\caption{Cell D2 after indentation. (a) Front face; scale bar, about 10 mm. (b) Back face.}\label{fig:S5}\end{minipage}
\end{figure}